\documentclass[11pt]{article}
 \usepackage{epsfig}
 \usepackage{color}
 \usepackage{cite}
 \def\be{\begin{equation}}
 \def\ee{\end{equation}}
 \def\bea{\begin{eqnarray}}
 \def\eea{\end{eqnarray}}
 \usepackage{graphicx}
 \def\simlt{\stackrel{<}{{}_\sim}}
\def\simgt{\stackrel{>}{{}_\sim}}

 \catcode`\@=11
 \def\lsim{\mathrel{\mathpalette\@versim<}}
 \def\gsim{\mathrel{\mathpalette\@versim>}}
 \def\@versim#1#2{\vcenter{\offinterlineskip
 \ialign{$\m@th#1\hfil##\hfil$\crcr#2\crcr\sim\crcr } }}
 \catcode`\@=12

 \catcode`@=12
\begin{document}
 \thispagestyle{empty}
 \begin{flushright}
 UCRHEP-T618\\
 December 2021\
 \end{flushright}
 \vspace{0.6in}
 \begin{center}
 {\LARGE \bf Predestined Dark Matter Varieties\\ 
 in the Simplest Left-Right Model\\}
 \vspace{0.8in}
 {\bf Talal Ahmed Chowdhury\\}
 \vspace{0.1in}
{\sl Department of Physics, University of Dhaka, P.O. Box 1000, 
Dhaka, Bangladesh\\}
\vspace{0.2in}
{\bf Shaaban Khalil\\}
 \vspace{0.1in}
{\sl Center for Fundamental Physics, Zewail City of Science and Technology,\\ 
6 October City, Giza 12588, Egypt\\}
\vspace{0.2in}
{\bf Ernest Ma\\}
 \vspace{0.1in}
{\sl Department of Physics and Astronomy,\\ 
University of California, Riverside, California 92521, USA\\}
\end{center}
 \vspace{0.8in}

\begin{abstract}\
In the simplest left-right extension of the Standard Model of particle 
interactions with one scalar bidoublet and one $SU(2)_R$ triplet for 
seesaw Majorana neutrino masses, the addition of a variety of fermion 
and scalar multiplets automatically makes them stable (predestined) 
dark matter candidates.  We discuss the interplay of this dark ensemble 
in relic abundance and direct searches.
\end{abstract}

 \section{Introduction}
To allow for massive neutrinos, the simplest way is to add right-handed 
neutrinos $\nu_R$ to the Standard Model (SM) of quarks and lepons.  However, 
since $\nu_R$ is a trivial singlet under the gauge symmetry 
$SU(3)_C \times SU(2)_L \times U(1)_Y$, its presence is rather {\it ad hoc}. 
To justify its existence, the left-right extension~\cite{LR1} is often considered, 
where the $\nu_R$ is now part of an $SU(2)_R$ doublet.  To break the gauge 
symmetry $SU(3)_C \times SU(2)_L \times SU(2)_R \times U(1)_{(B-L)/2}$ to 
$SU(3)_C \times U(1)_Q$, there are several options~\cite{m04}, the simplest 
is to have just one $SU(2)_L \times SU(2)_R$ scalar bidoublet which provides 
masses for all fermions, including a Dirac mass linking $\nu_L$ and $\nu_R$, 
and one $SU(2)_R \times U(1)_{(B-L)/2}$ scalar triplet for breaking the latter 
to $U(1)_Y$ at a high scale and endowing $\nu_R$ with a large Majorana mass. 
The well-known seesaw mechanism then ensures that $\nu_L$ gets a naturally 
small Majorana mass.

There is of course no dark-matter candidate~\cite{bh18} so far in this 
discussion.  Suppose a new fermion or scalar multiplet is added to 
the above minimal particle content.  It may in fact become suitable dark 
matter, because the new theory (including all its interactions with the 
existing particles) now has an extra conserved symmetry which makes the 
lightest particle of this multiplet stable.  This notion of ``predestined'' 
dark matter~\cite{m18} has recently been proposed.  It is the extension 
of previous work~\cite{cfs06} based on the SM.  In this paper, various 
examples are studied in some detail to show how they fit the requirements 
of relic abundance and direct-search constraints.

 \section{Left-Right Model with Dark Multiplets}\label{DMLRsec}
 
Under $SU(3)_C \times SU(2)_L \times SU(2)_R \times U(1)_{(B-L)/2}$, 
with quarks and leptons transforming as
\begin{eqnarray}
&& (u,d)_L \sim (3,2,1,1/6), ~~~ (u,d)_R \sim (3,1,2,1/6), \\
&& (\nu,l)_L \sim (1,2,1,-1/2), ~~~ (\nu,l)_R \sim (1,1,2,-1/2),
\end{eqnarray}
a scalar bidoublet
\begin{equation}
\eta = \pmatrix{\eta_1^0 & \eta_2^+ \cr \eta_1^- & \eta_2^0} \sim (1,2,2,0) 
\end{equation}
with $\langle \eta_{1,2}^0 \rangle \neq 0$ links them to form Dirac fermion 
masses.  An $SU(2)_R \times U(1)_{(B-L)/2}$ triplet
\begin{equation}
\xi_R = (\xi_R^{++},\xi_R^+,\xi_R^0) \sim (1,1,3,1)
\end{equation}
then provides the proper symmetry breaking to $U(1)_Y$ and allows $\nu_R$ 
to have a large Majorana mass.
 
In this framework, it has been shown~\cite{m18} that the following simple 
multiplets are automatic (predestined) dark-matter cnadidates, i.e. the 
addition of any one of them by itself to the Lagrangian of this left-right 
model brings with it an unbroken $Z_2$ symmetry, without being imposed, which 
would guarantee its stability.  There are four fermion multiplets
\begin{eqnarray}
&& S \sim (1,1,1,0); ~~~ \psi = \pmatrix{\psi_1^0 & \psi_2^+ \cr 
\psi_1^- & \psi_2^0} \sim (1,2,2,0); \\ 
&& \Sigma_L = (\Sigma_L^+, \Sigma_L^0, \Sigma_L^-) \sim (1,3,1,0); ~~~ 
\Sigma_R = (\Sigma_R^+, \Sigma_R^0, \Sigma_R^-) \sim (1,1,3,0).
\end{eqnarray}
Individually, they have mass terms
\begin{eqnarray}
&& {1 \over 2} m_S SS; ~~~ {1 \over 2} m_\psi Tr (\psi \tilde{\psi}^\dagger) 
= \psi_1^0 \psi_2^0 - \psi_1^- \psi_2^+ = det(\psi); \\ 
&& {1 \over 2} m_L (\vec{\Sigma}_L \cdot \vec{\Sigma}_L); ~~~  
{1 \over 2} m_R (\vec{\Sigma}_R \cdot \vec{\Sigma}_R),
\end{eqnarray}
where
$\tilde{\psi} = \sigma_2 \psi^* \sigma_2$.  Except for $S$, they all have 
gauge interactions, but the fermions $\psi,\Sigma_{L,R}$ have no contact 
with the quarks and leptons through the available scalar bosons $\eta,\xi_R$. 
Hence a symmetry exists (which is in fact $Z_2$ as evidenced 
from the mass terms above) to protect the lightest memeber of each multiplet 
from decaying.

All four multiplets may be linked by inserting a real singlet scalar 
$\zeta \sim (1,1,1,0)$, which acts as a mediator among the four separate 
dark sectors and that of the particles of the left-right model. 
However, without imposing any symmetry, there will be one more term
\begin{equation}
S Tr(\psi \tilde{\eta}^\dagger).
\end{equation}
This shows that there are only three separate dark sectors, with the 
symmetry $Z_2 \times Z_2 \times Z_2$.  Here we consider the following 
two instructive cases, each with only one dark sector. 
\begin{itemize}
\item{ (1) Only $\Sigma_R$ is added.  The dark symmetry is $Z_2$ and the 
Majorana fermion $\Sigma^0_R$ is dark matter.}
\item{ (2) Only $S$ and $\psi$ are added.  Then both share the same $Z_2$, 
under which both are odd.  The lightest of the three mass eigenstates 
spanning $(S,\psi^0_1,\psi^0_2)$ is dark matter.}
\end{itemize}

\section{Gauge Bosons in Left-Right Model}\label{gaugebosonLR}

In this section, we present the notations used to describe the gauge boson sector of the Left-Right Model. Let $\langle \eta^0_{1,2} \rangle = v_{1,2}$ and $\langle \xi_R^0 \rangle = v_R$, 
then the $2 \times 2$ mass-squared matrix spanning $(W_L^\pm,W_R^\pm)$ is 
\begin{equation}
{\cal M}^2_W = \pmatrix {(1/2)g_L^2(v_1^2+v_2^2) & -g_L g_R v_1 v_2 \cr 
-g_L g_R v_1 v_2 & (1/2)g_R^2 (v_1^2+v_2^2+2v_R^2)},
\end{equation}
and the $3 \times 3$ mass-squared matrix spanning $(W_L^0,W_R^0,B)$ is
\begin{equation}
{\cal M}^2_Z = \pmatrix {(1/2) g_L^2(v_1^2+v_2^2) & -(1/2)g_L g_R (v_1^2+v_2^2)
 & 0 \cr -(1/2) g_L g_R (v_1^2+v_2^2) & (1/2)g_R^2(v_1^2+v_2^2+4v_R^2) & 
-2g_R g_B v_R^2 \cr 0 & -2g_R g_B v_R^2 & 2g_B^2 v_R^2}.
\end{equation}
Using $e^{-2} = g_L^{-2} + g_R^{-2} + g_B^{-2}$, and assuming $g_L=g_R$ with 
$x=\sin^2 \theta_W$, the physical neutral gauge bosons are
\begin{equation}
\pmatrix{A \cr Z \cr Z'} = \pmatrix{\sqrt{x} & \sqrt{x} & \sqrt{1-2x} \cr 
\sqrt{1-x} & -x/\sqrt{1-x} & -\sqrt{x(1-2x)/(1-x)} \cr 0 & 
\sqrt{(1-2x)/(1-x)} & -\sqrt{x/(1-x)} } \pmatrix{W_L^0 \cr W_R^0 \cr B}.
\label{neutralgaugeboson}
\end{equation}
The photon $A$ is massless.  The $2 \times 2$ mass-squared matrix spanning 
$(Z,Z')$ is
\begin{equation}
{\cal M}^2_{Z,Z'} = {e^2 \over 2} \pmatrix{ (v_1^2+v_2^2)/x(1-x) & 
-\sqrt{1-2x} (v_1^2+v_2^2)/x(1-x) \cr -\sqrt{1-2x} (v_1^2+v_2^2)/x(1-x) 
& 4(1-x)v_R^2/x(1-2x) + (1-2x)(v_1^2+v_2^2)/x(1-x)}.
\end{equation}

Moreover, the coupling of $Z$ and $Z'$ with the LH and RH fermion fields $f_{L,R}$ are denoted as $g_{Z}(f_{L,R})$ and $g_{Z'}(f_{L,R})$, respectively, and with $g_{L}=g_{R}=g$ they are expressed as follows,
\begin{equation}
g_{Z}(f_{L})=\frac{g}{\sqrt{1-x}}\left[T_{3L}(f_{L})-x Q(f_{L})\right],\,\,\, g_{Z}(f_{R})=\frac{-g\,x}{\sqrt{1-x}} Q(f_{R}),\nonumber
\end{equation}
\begin{eqnarray}
g_{Z'}(f_{L})&=&\frac{g}{\sqrt{(1-x)(1-2x)}}\left[T_{3L}(f_{L}) - Q(f_{L})\right],\\
g_{Z'}(f_{R})&=&\frac{g}{\sqrt{(1-x)(1-2x)}}\left[(1-x)T_{3R}(f_{R})-x Q(f_{R})\right].
\end{eqnarray}
where, $Q$ is the electric charge operator acting on the fermion fields.

\section{$SU(2)_R$ Fermion Triplet Dark Matter}\label{sigmaR-sec} 

Consider the $SU(2)_R$ fermion triplet $\Sigma_R$ by itself.  It interacts 
with only the $SU(2)_R$ gauge bosons $W_R^\pm,W_R^0$.  However, since 
$\Sigma_R^-$ is identical to the conjugate of $\Sigma_L^+$, the gauge 
interactions of $\Sigma_R$ are
\begin{equation}
g_R {W_R^0}_\mu \overline{\Sigma^+} \gamma^\mu \Sigma^+ 
+ g_R {W_R^+}_\mu \overline{\Sigma^+} \gamma^\mu  \Sigma^0
+ g_R {W_R^-}_\mu \overline{\Sigma^0} \gamma^\mu  \Sigma^+.\nonumber
\end{equation}
In the above, the subscript $R$ has been dropped from $\Sigma$ to avoid the 
confusion with the chirality projection operator.  However, the physical 
particles themselves will still be denoted as $\Sigma_R^\pm,\Sigma_R^0$. 
They have the same invariant mass $m_\Sigma$ at tree level, but the 
one-loop exchange of $W_R^\pm,W_R^0$ as shown in Figs.~1,2 will make 
$\Sigma_R^\pm$ heavier~\cite{s95} than $\Sigma^0_R$, which is then a 
dark-matter candidate.
\begin{figure}[htb]
\vspace*{-5cm}
\hspace*{-3cm}
\includegraphics[scale=1.0]{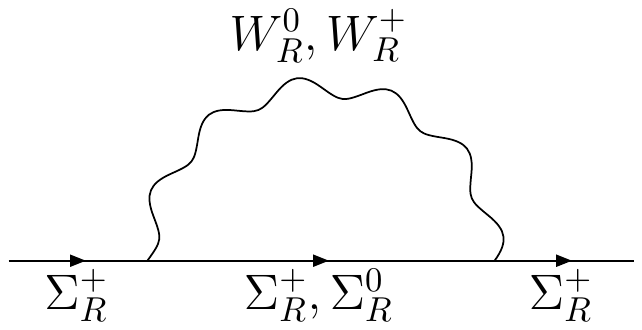}
\vspace*{-21.5cm}
\caption{One-loop correction to $\Sigma_R^+$ mass.}
\label{one-loop1}
\end{figure}
Whereas $W_R^\pm$ is mostly a mass eigenstate for $v_1 v_2 << v_R^2$, 
$W_R^0$ is a linear combination of $A,Z$ and $Z'$. 
\begin{figure}[htb]
\vspace*{-5cm}
\hspace*{-3cm}
\includegraphics[scale=1.0]{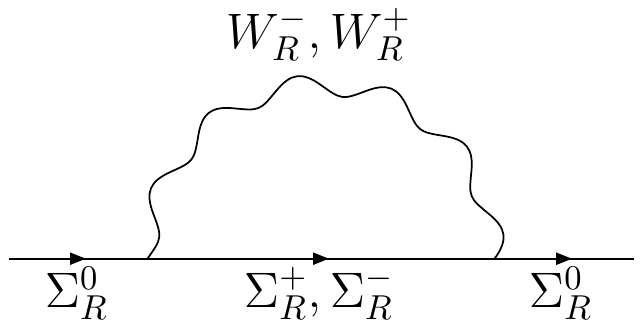}
\vspace*{-21.5cm}
\caption{One-loop correction to $\Sigma_R^0$ mass.}
\label{one-loop2}
\end{figure}
Each of the two loop contributions is infinite, but the 
difference is finite, i.e.
\begin{equation}
m(\Sigma_R^\pm)-m(\Sigma_R^0) = {\alpha m_\Sigma \over 2 \pi x} \left[ 
f\left( {m^{2}_{W_R} \over m^{2}_\Sigma} \right) - {1-2x \over 1-x} f \left( 
{m^{2}_{Z'} \over m^{2}_\Sigma} \right) - {x^2 \over 1-x} 
f \left( {m^{2}_Z \over m^{2}_\Sigma} \right) \right],
\label{massspliteq}
\end{equation}
where
\begin{equation}
f(r)=\int_{0}^{1} d x\,(1+x)\log\left[1+\frac{r(1-x)}{x^{2}}\right]\nonumber
\end{equation}

\begin{figure}[h!]
\centerline{\includegraphics[width=0.65\textwidth]{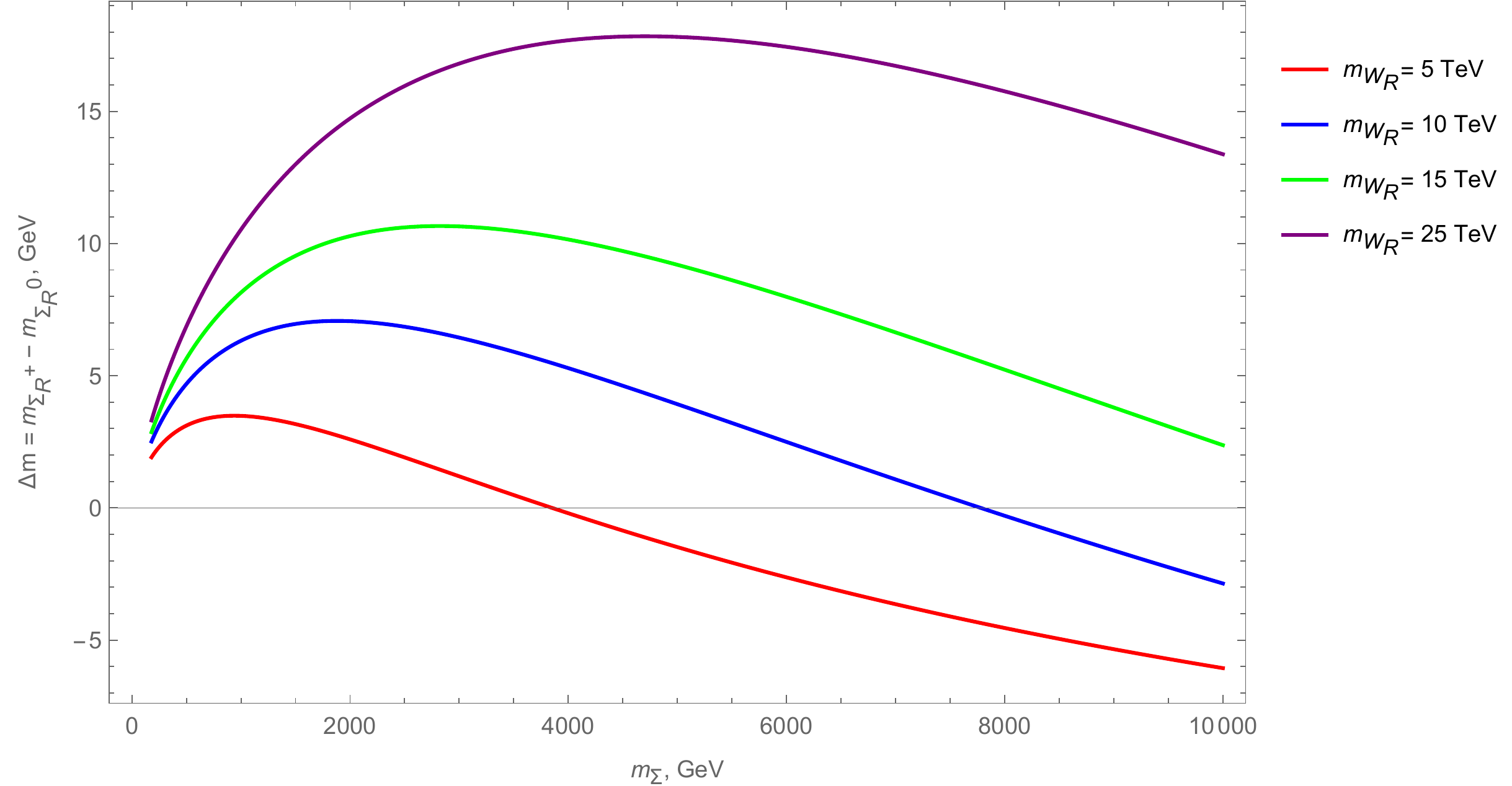}}
\caption{The mass splitting, $\Delta m$ between $\Sigma^{+}_{R}$ and $\Sigma^{0}_{R}$ at one-loop induced by gauge bosons as shown in Fig. \ref{one-loop1} and \ref{one-loop2}. Here we can see that, if $m_{\Sigma}$ is sufficiently large compared to $m_{W_{R}}$, we see that the neutral component, $\Sigma^{0}_{R}$ becomes heavier at one point, and ceases to be the dark matter candidate.}
\label{mass-split}
\end{figure}
We can see from Fig. \ref{mass-split} that if $m_{\Sigma}$ is sufficiently large enough compared to $m_{W_{R}}$, the charged component of $\Sigma_{R}$ becomes lighter than its neutral component, and the $\Sigma^{0}_{R}$ ceases to be a good dark matter candidate. Unlike the case of the SM where symmetry breaking is due to the vacuum expectation value (VEV) of the Higgs doublet of the $SU(2)_{L}$, here the $SU(2)_{R}$ breaking is given by the VEV of the scalar triplet $\xi_{R}$ which leads to $m^2_{Z'}/m^{2}_{W_{R}}\simeq 2(1-x)/(1-2x)$. As the function $f(r)\rightarrow 0$ as $r\rightarrow 0$, the contribution coming from the Z boson in the third term of Eq. \ref{massspliteq} is negligible. Therefore, if we set $m_{W_{R}}$ to a specific value, after some value of $m_{\Sigma}\simlt m_{W_{R}}$ the second term starts to become larger than the first term, and the sign of the mass splitting becomes negative. This means that to ensure the neutral component field, $\Sigma^{0}_{R}$ to be the lightest DM candidate, the parameter space has to satisfy $m_{\Sigma}<m_{W_{R}}$.  Besides, the mass gap is calculated to be of order 1 - 10 GeV for $m_{W_R}$ a few TeV as shown in Fig. \ref{mass-split}, so the decay $\Sigma^{\pm}_{R}\rightarrow \Sigma^{0}_{R}\,\pi^{\pm}$ is kinematically accessible via an off-shell emission of $W^{\pm}_{R}$, and the decay-width is approximately (when the momentum transfer $Q^{2}\ll m^{2}_{W_{R}}$) given as
\begin{equation}
\Gamma_{\Sigma^{\pm}_{R}\rightarrow \Sigma^{0}_{R}\,\pi^{\pm}}\sim \frac{g^4 \Delta m^{3}f_{\pi}^2}{m_{W_{R}}^{4}\pi}\sqrt{1-\frac{m_{\pi}^2}{\Delta m^{2}}}
\label{decaywidth}
\end{equation}
where $f_{\pi}$ is the pion decay constant. Now considering the $100\%$ decay of $\Sigma^{\pm}_{R}$ to pion channel, one can constrain the possible value of $(m_{\Sigma},\,m_{W_{R}})$ from searches of disappearing tracks at the LHC.

As $W_R^\pm$ has to be heavier than $\Sigma_R^0$, then the relevant 
annihilation channels are $\Sigma_R^0 \Sigma_R^\pm \to W_R^\pm \to $ SM 
particles, and $\Sigma_R^+ \Sigma_R^- \to (A,Z,Z') \to $ SM particles.
Assuming that $\nu_R$ is too heavy to be produced, and taking into 
account 3 quark colors and 3 families, the first process yields
\begin{equation}
\sigma^{(1)}_{ann} \times v_{rel} = {36 \pi \alpha^2 m_\Sigma^2 \over x^2 
[4m^2_\Sigma - m^2_{W_R}]^2},
\label{sigmaW}
\end{equation}
and the second yields
\begin{eqnarray}
\sigma^{(2)}_{ann} \times v_{rel} &=& { \pi \alpha^2 m_\Sigma^2 \over 
(1-x)^2} \left\{ 42 \left[ {1 \over 4m^2_\Sigma} + {1-2x \over x(4m^2_\Sigma - 
m^2_{Z'})}\right]^2 + 20 \left[ {1 \over 4m^2_\Sigma} - {1 \over 4m^2_\Sigma 
- m^2_{Z'}} \right]^2 \right. \nonumber \\
&+& \left. 12 \left[ {1 \over 4m^2_\Sigma} + {1-2x \over x(4m^2_\Sigma - 
m^2_{Z'})}\right] \left[ {1 \over 4m^2_\Sigma} - {1 \over 4m^2_\Sigma 
- m^2_{Z'}} \right] + {1 \over m^4_\Sigma} \right\}
\label{sigmaZ}
\end{eqnarray}
where $m^2_Z$ has been neglected against $4m^2_\Sigma$, and the last term is 
the $AA, AZ, ZZ$ contribution.

\begin{figure}[h!]
\centerline{\includegraphics[width=.65\textwidth]{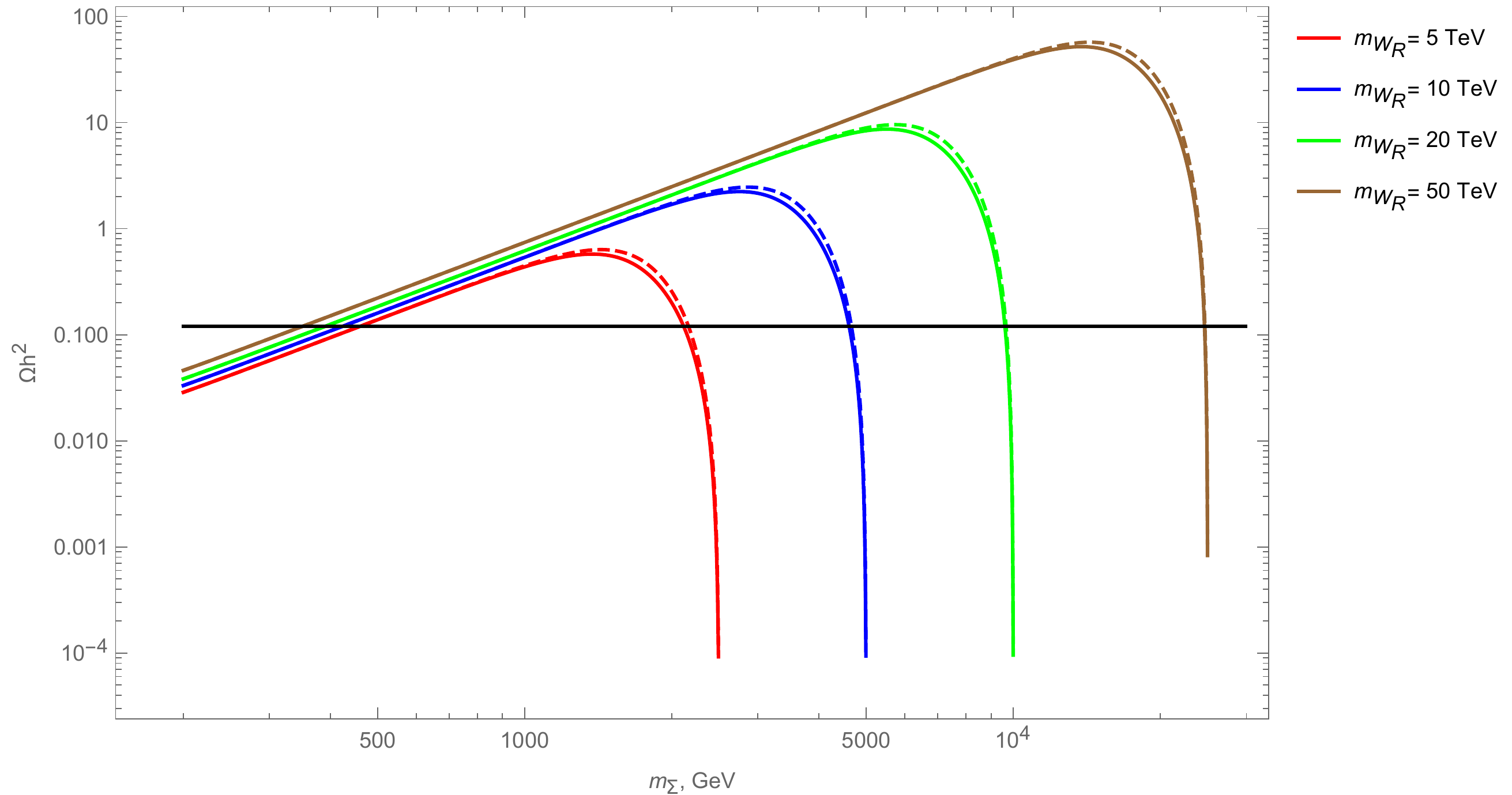}\hspace{0mm}\includegraphics[width=.5\textwidth]{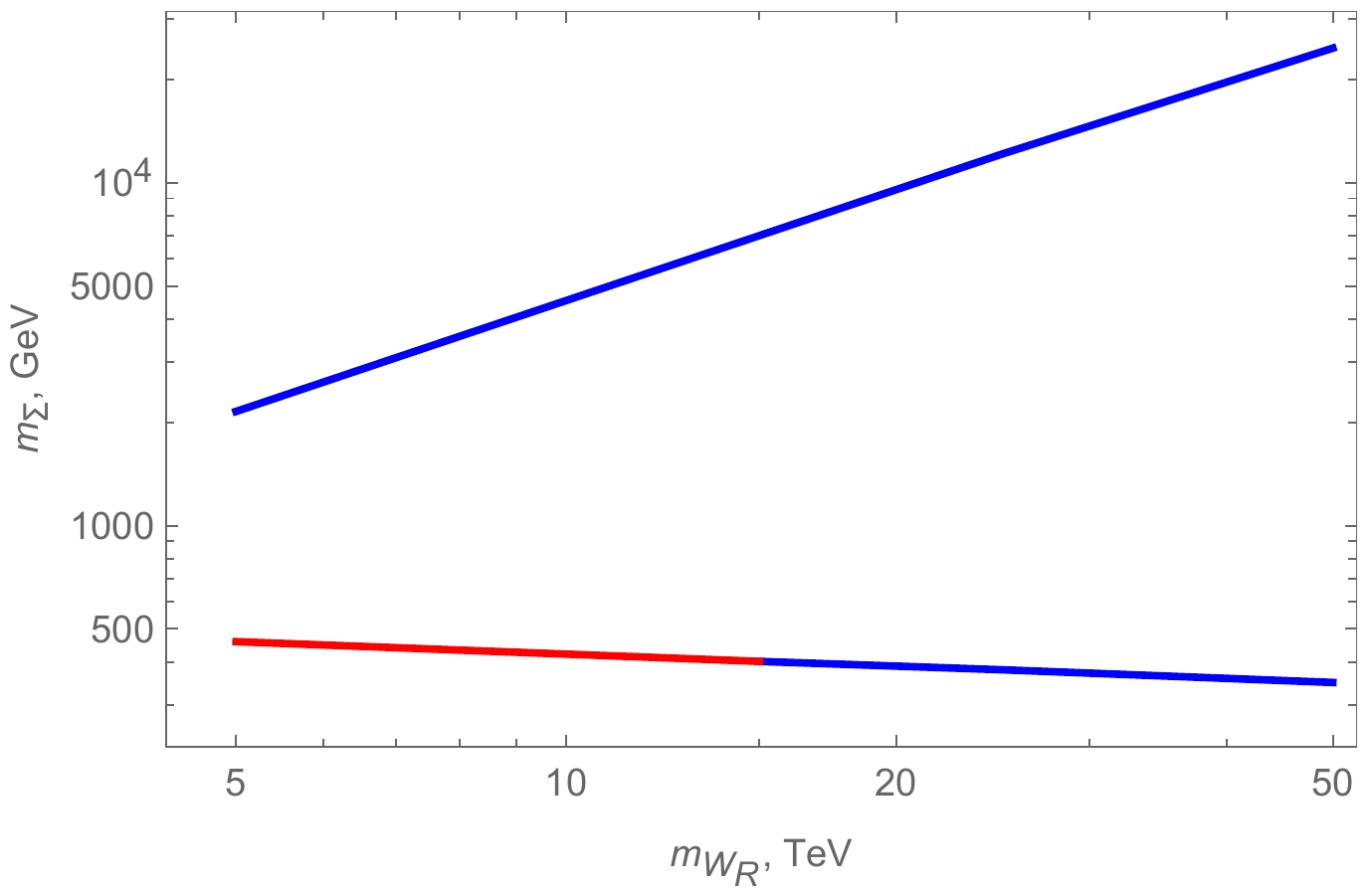}}
\caption{(Left) The DM relic density with respect to the DM mass, $m_{\Sigma}$ for different values of $m_{W_{R}}$. The black line represents the central value of the observed DM relic density, $\Omega h^{2}=0.12\pm 0.001$ \cite{Aghanim:2018eyx}. Besides, the solid red, blue, green and brown lines represent the relic densities for RH neutrino masses, $m_{N_{1,2,3}}=100$ GeV, and the dashed lines represent the relic densities for RH neutrino masses, $m_{N_{1,2,3}}=m_{W_{R}}/2$, respectively. We can see that the value of RH neutrino masses make very small impact in the relic density calculation.  (Right) The correlation between $m_{W_{R}}$ and $m_{\Sigma}$ satisfying the correct DM relic abundance. Here the red color indicates the values of $(m_{W_{R}},\,m_{\Sigma})$ excluded by the null searches of disappearing tracks at the CMS \cite{CMS:2020atg}.}
\label{relicplots}
\end{figure}

Now if we consider the standard thermal freeze-out via the above-mentioned coannihilation channels to achieve the DM relic abundance of the universe, for a fixed value of $m_{W_{R}}$, the correct relic abundance \cite{Aghanim:2018eyx} is given by two values of DM mass, $m_{\Sigma}$ as seen from Fig.~\ref{relicplots} (left).  As the $m_{W_R}$ increases from its lower  bound~\cite{atlas19} of 5 TeV to larger values, the lower branch of the DM mass  $m_{\Sigma}$ set by the dominant contribution from $\sigma^{(2)}_{ann}$, slowly decreases from the value of $m_{\Sigma}= 402$ GeV (for  $m_{W_{R}} = 50$ TeV, $m_{\Sigma}$ is $349$ GeV). On the other hand, if $\sigma^{(1)}_{ann}$ contributes the most, then $m_\Sigma = 2.15$ TeV for $m_{W_R} = 5$ TeV and increases as the latter increases (for  $m_{W_{R}} = 50$ TeV, $m_{\Sigma}$ is $24.6$ TeV). We plot this behavior in Fig.~\ref{relicplots} (right). It is worth mentioning that our result matches with the finding of \cite{Heeck:2015qra}, where the $SU(2)_{R}$ fermion triplet DM was studied in the context of minimal Left-Right Symmetric Model.

\section{Fermion singlet DM via Freeze-in mechanism}\label{singletS-sec}

In this section we consider the second possibility mentioned in section \ref{DMLRsec} where fermion singlet $S$ and bidoublet $\psi$ are added to the matter content, both of them being odd under the same $Z_2$. The interaction between $S$ and $\psi$ is given by the following Yukawa term
\begin{equation}
f_S S Tr(\psi \tilde{\eta}^\dagger) = f_S S(\psi_1^0 \eta_2^0 + \psi_2^0 \eta_1^0 
- \psi_1^- \eta_2^+ - \psi_2^+ \eta_1^-).
\label{sint1}
\end{equation}
When the coupling $|f_{S}|\sim O(1)$, the neutral component fields, $S$ and $\psi^{0}_{1,2}$ will have non-negligible mixing induced by this term Eq. \ref{sint1} after symmetry breaking, and the lightest Majorana mass-eigenstate of them will be the DM candidate. In this case, the standard thermal freeze-out mechanism will set the relic abundance of the DM candidate which has been studied in \cite{Berlin:2016eem}.

Instead, we explore the possibility of $S$ being the DM candidate whose relic abundance is achieved via freeze-in mechanism \cite{McDonald:2001vt}. The component fields of the fermionic bidoublet, $\psi$ and scalar bidoublet, $\eta$, having gauge interactions, act as the mediator or bath particles whose decay and scattering would create the abundance of $S$.

In our set-up, as the coupling $f_S$ controls the freeze-in, and has to be of the order $O(10^{-10}-10^{-9})$ (for review, please see \cite{Bernal:2017kxu}), we neglect the mixing between $S$ and $\psi^{0}_{1,2}$. As $\psi^{0}_{1}$ and $\psi^{0}_{2}$ have opposite $W^{3}_{L,R}$ charges, we define the corresponding Majorana spinors,
\begin{equation}
\chi_{1}=\frac{1}{\sqrt{2}}\left(\psi^0_{1}+\psi^{0}_{2}\right),\,\,\,\chi_{2}=\frac{1}{\sqrt{2} i}\left(\psi^0_{1}-\psi^{0}_{2}\right)
\label{majorana}
\end{equation}
Also, the charged component fields are collected in the corresponding 4-component spinor, $\Psi^{-}=\left(\psi^{-}_{1},\psi^{-}_{2}\right)^{T}$.
We also neglect the radiative corrections between the charged and neutral components of the fermion bidoublet because we are not considering the lightest neutral component field of $\psi$ as the DM candidate, and hence $\Psi^{\pm}$ and $\chi_{1,2}$ have the common tree-level mass, $m_{\psi}$. Now the corresponding gauge interactions relevant for $\Psi^{\pm}$ and $\chi_{1,2}$ are,
\begin{eqnarray}
{\cal L}_{\mathrm{charged-current}}& \supset & \frac{g}{2}\left[\overline{\chi_{1}}\gamma^{\mu}\Psi^{-}W^{+}_{L\mu}-i\, \overline{\chi_{2}}\gamma^{\mu}\Psi^{-}W^{+}_{L\mu}+\overline{\chi_{1}}\gamma^{\mu}\Psi^{-}W^{+}_{R\mu}-i\, \overline{\chi_{2}}\gamma^{\mu}\Psi^{-}W^{+}_{R\mu}\right]+\mathrm{h.c}\label{gaugeint1}\\
{\cal L}_{\mathrm{neutral-current}}&\supset & e \overline{\Psi^{-}}\gamma^{\mu}\Psi^{-}+\frac{g}{2}\,\left[(1-2x)/\sqrt{1-x}\,\,\overline{\Psi^{-}}\gamma^{\mu}\Psi^{-}Z_{\mu}+\,\sqrt{(1-2x)/(1-x)}\,\,\overline{\Psi^{-}}\gamma^{\mu}\Psi^{-}Z^{'}_{\mu}\right]\nonumber\\
&+&i\,\frac{g}{2}\,\left[1/\sqrt{1-x}\,\,\overline{\chi_{2}}\gamma^{\mu}\chi_{1}Z_{\mu}-\sqrt{(1-2x)/(1-x)}\,\,\overline{\chi_{2}}\gamma^{\mu}\chi_{1}Z^{'}_{\mu}\right]+\mathrm{h.c}\label{gaugeint2}
\end{eqnarray}
Besides, the Yukawa term of Eq. \ref{sint1} leads to,
\begin{equation}
{\cal L}_{Y}\supset \frac{f_{S}}{\sqrt{2}}\left[\overline{S}P_{L}\,\chi_{1}\left(\eta_{1}^{0}+\eta^{0}_{2}\right)+i\, \overline{S}P_{L}\,\chi_{2}\left(\eta^{0}_{2}-\eta^{0}_{1}\right)-\overline{S}P_{L}\,\Psi^{-}\eta^{+}_{2}+\overline{\Psi^{-}}P_{L}\,S\eta^{-}_{1}\right]+\mathrm{h.c}
\label{sint2}
\end{equation}
where, the yukawa coupling, $f_{S}=a+ i\, b$.
The component fields of the scalar bidoublet are expressed in terms of the mass-eigenstates in the limit, $v_{1,2}\ll v_{R}$ as,
\begin{eqnarray}
\eta^{0}_{1}+\eta^{0}_{2}&\sim &\frac{1}{\sqrt{2}}\left(\cos\beta+\sin\beta\right) h +\frac{1}{\sqrt{2}}\left(\cos\beta-\sin\beta\right)H-\frac{i}{\sqrt{2}}\left(\cos\beta+\sin\beta\right)A^{0}\nonumber\\
\eta^{0}_{2}-\eta^{0}_{1}&\sim & \frac{1}{\sqrt{2}}\left(\sin\beta-\cos\beta\right) h+\frac{1}{\sqrt{2}}\left(\cos\beta+\sin\beta\right) H^{0} -\frac{i}{\sqrt{2}}\left(\cos\beta-\sin\beta\right)A^{0}\nonumber\\
\eta^{+}_{1}&\sim & \cos\beta\,\, H^{+},\,\,\,\eta^{+}_{2}\sim \sin\beta\,\, H^{+}
\end{eqnarray}
where, $\tan\beta=v_{2}/v_{1}$.

\subsection{$S$ as the Freeze-in DM}
The freeze-in mechanism of the singlet fermion, $S$ is controlled by the Yukawa term, Eq. \ref{sint2} via the decay and scattering of the bath particles which are the fermions, $\Psi^{\pm}$, $\chi_{1,2}$ (with degenerate mass, $m_{\psi}$), and the scalars, $h$, $H^{0}$, $A^{0}$ and $H^{\pm}$. The observational constraints from flavor changing neutral currents (FCNC) set the masses of heavy Higgses $(H^{0},\,A^{0},\,H^{\pm})$ to be $m_{H}\simgt 20$ TeV \cite{Bertolini:2014sua}. Depending on the mass values, there are two possible classes of decay channels which would produce $S$ out-of-equilibrium; I. $m_{H}>m_{\psi}+m_{S}$ and II. $m_{\psi}>m_{H}+m_{S}$. As $S$ is the DM candidate, $m_{S}<m_{H},m_{\psi}$. Now the decay channels are,
\begin{equation}
h\rightarrow \chi_{1}\,\,S,\,\,\,H^{0}\rightarrow \chi_{2}\,\, S,\,\,\, A\rightarrow \chi_{1,2}\,\,S,\,\,\,H^{\pm}\rightarrow \Psi^{\pm}\,\,S\,\,\,\mathrm{when}\,\,\,m_{H, h}>m_{\psi}+m_{S},
\label{decay1}
\end{equation}
and
\begin{equation}
\chi_{1}\rightarrow h\,\,S,\,\,\,\chi_{2}\rightarrow H^{0}\,\,S,\,\,\,\chi_{1,2}\rightarrow A^{0}\,\,S,\,\,\,\Psi^{\pm}\rightarrow H^{\pm}\,\,S\,\,\,\mathrm{when}\,\,\,m_{\psi}>m_{H, h}+m_{S}
\label{decay2}
\end{equation}
In the subsequent analysis, we will consider $m_{\psi}\gg m_{h}$ so the decay channel, $h\rightarrow \chi_{1}\,\,S$ is forbidden.

\begin{table}
\begin{tabular}{|c|c|c|c|}
\hline
Initial state(s) & Final state(s) & Feynman Diagrams & {\scriptsize Amplitude's coupling order} \\
\hline
$\chi_{i}\,\chi_{j}\,\,(i,j=1,2)$ & $S\,S$ & $t(h,H^{0},A^{0})$, $u(h,H^{0},A^{0})$ & $O(f_{S}^{2})$\\
$\Psi^{\pm}\,\Psi^{\mp}$ & $S\,S$ & $t(H^{\pm})$, $u(H^{\pm})$ & \\
$h h,\,H^{0} H^{0},\,A^{0} A^{0},\,h H^{0},\,h A^{0},\, H^{0}A^{0}$ & $S\,S$ & $t(\chi_{i})$, $u(\chi_{i})$, $(i=1,2)$ & \\
$H^{\pm}\,H^{\mp}$ & $S\,S$ & $t(\Psi^{\pm})$, $u(\Psi^{\pm})$ & \\
\hline
$\chi_{i}\,Z$, $\chi_{i}\,Z^{'}$ & $h\,S$, $H^{0}\,S$, $A^{0}\,S$ & $s(\chi_{j})$ & $O(g\,f_{S})$ \\
$\Psi^{\pm}\gamma$, $\Psi^{\pm}\,Z$, $\Psi^{\pm}\,Z^{'}$ & $H^{\pm}\,S$ & $s(\Psi^{\pm})$ & \\
$h Z$, $h Z^{'}$, $H^{0} Z$, $H^{0} Z^{'}$ & $\chi_{i}\, S,\,(i=1,2)$ & $s(A^{0})$, $t(\chi_{j})\, (j\neq i)$ & \\
$A^{0} Z$, $A^{0} Z^{'}$ & $\chi_{i}\, S,\,(i=1,2)$ & $s(h, H^{0})$, $t(\chi_{j})$ & \\
$H^{\pm}\gamma,\,H^{\pm} Z,\,H^{\pm} Z^{'}$ & $\Psi^{\pm}\,S$ & $s(H^{\pm})$, $t(\Psi^{\pm})$ & \\
$H^{\pm} W^{\mp}_{L,R}$ & $\chi_{i}\,S$ & $s(h, H^{0}, A^{0})$ & \\
$f\,\overline{f}$ & $\chi_{i}\,S$ & $s(h, H^{0}, A^{0})$ & \\
$W^{\pm}_{L,R}W^{\mp}_{L,R}$, $Z Z$, $Z^{'}Z^{'}$ & $\chi_{i}\,S$ &  $s(h, H^{0}, A^{0})$ & \\
$h \chi_{i}$, $H^{0}\chi_{i}$, $ A^{0}\chi_{i}$ & $Z S$, $Z^{'} S$ & $t(\chi_{j})$, $(j\neq i)$ & \\
$H^{\pm}\Psi^{\mp}$ & $\gamma S,\, Z S,\,Z^{'} S$ & $t(\Psi^{\pm})$ & \\
\hline
\end{tabular}
\caption{ Possible scattering channels between different bath particles which lead to the production of $S$. Here $s(x)$, $t(x)$ and $u(x)$ represent the $``x"$ particle in the s-channel, t-channel and u-channel, respectively.}
\label{scatteringtab}
\end{table}

Apart from the decay, $S$ is also produced through the scattering of bath particles. We tabulate the possible scattering channels in Table \ref{scatteringtab} in which we can see that most of the scattering channels are suppressed either by the large masses ($m_{\psi}\sim O(\mathrm{TeV}), m_{H}\sim 20$ TeV) in the propagators and smaller initial number densities as $\Psi^{\pm},\chi_{1,2}, H^{0},A^{0},H^{\pm},W^{\pm}_{R},Z^{'}$ are all non-relativistic with masses in the TeV range. Therefore, only the channels $f\overline{f},\,W^{+}_{L}W^{-}_{L},\,Z Z\rightarrow \chi_{1,2}\,S$ with $h$ in the s-channel will be less suppressed. For this reason, we consider these two channels for our calculation of relic abundance via freeze-in mechanism.
The yield of the FIMP DM candidate, $S$ at present temperature of the universe, $T_{0}$, denoted by $Y^{0}_{S}$, is given as \cite{Belanger:2018ccd}
\begin{equation}
Y^{0}_{S}=\int^{T_{R}}_{T_{0}}\frac{d T}{T \overline{H}(T) s(T)}\left({\cal N}(x\rightarrow S\,y)+{\cal N}(x\, x'\rightarrow S\, y)\right)
\label{yield}
\end{equation}
Here, ${\cal N}(x\rightarrow S\,y)$ and ${\cal N}(x\, x'\rightarrow S\,y)$ represent thermal collision terms corresponding to the decay channels given in Eq. \ref{decay1} or \ref{decay2}, and scattering channels given in Table \ref{scatteringtab}, respectively. In the radiation dominated universe, the energy and entropy densities are given by,
\begin{equation}
\rho(T)=\frac{\pi^{2}}{30}g_{eff}(T)T^{4},\,\,\,s(T)=\frac{2\pi^{2}}{45}h_{eff}(T)T^{3}
\label{densities}
\end{equation}
where, $g_{eff}$ and $h_{eff}$ are the total and the relativistic number of degrees of freedom, respectively.
The Hubble expansion rate, $H(T)$ and the modified Hubble rate, $\overline{H}(T)$ used in Eq. \ref{yield} are given as,
\begin{equation}
H(T)=\frac{1}{M_{pl}}\sqrt{\frac{8\pi}{3}\rho(T)},\,\,\,\mathrm{and}\,\,\,\overline{H}(T)=\frac{H(T)}{1+\frac{1}{3}\frac{d\, \ln(h_{eff}(T))}{d\, \ln T}}.
\label{hubble}
\end{equation}
Finally, the relic density of $S$ is given by,
\begin{equation}
\Omega h^{2}=\frac{m_{S} Y^{0}_{S} s_{0} h^{2}}{\rho_{c}}
\label{relic}
\end{equation}
where, the present entropy density of the universe is $s_{0}=2891.2\,\mathrm{cm}^{-3}$ and the critical density, $\rho_{c}=1.053 \times 10^{-5}\, h^{2}\,\mathrm{GeV cm}^{-3}$, and the scaling factor for Hubble expansion rate, $h=0.674$ \cite{Zyla:2020zbs}.
\begin{figure}[h!]
\centerline{\includegraphics[width=9.5cm]{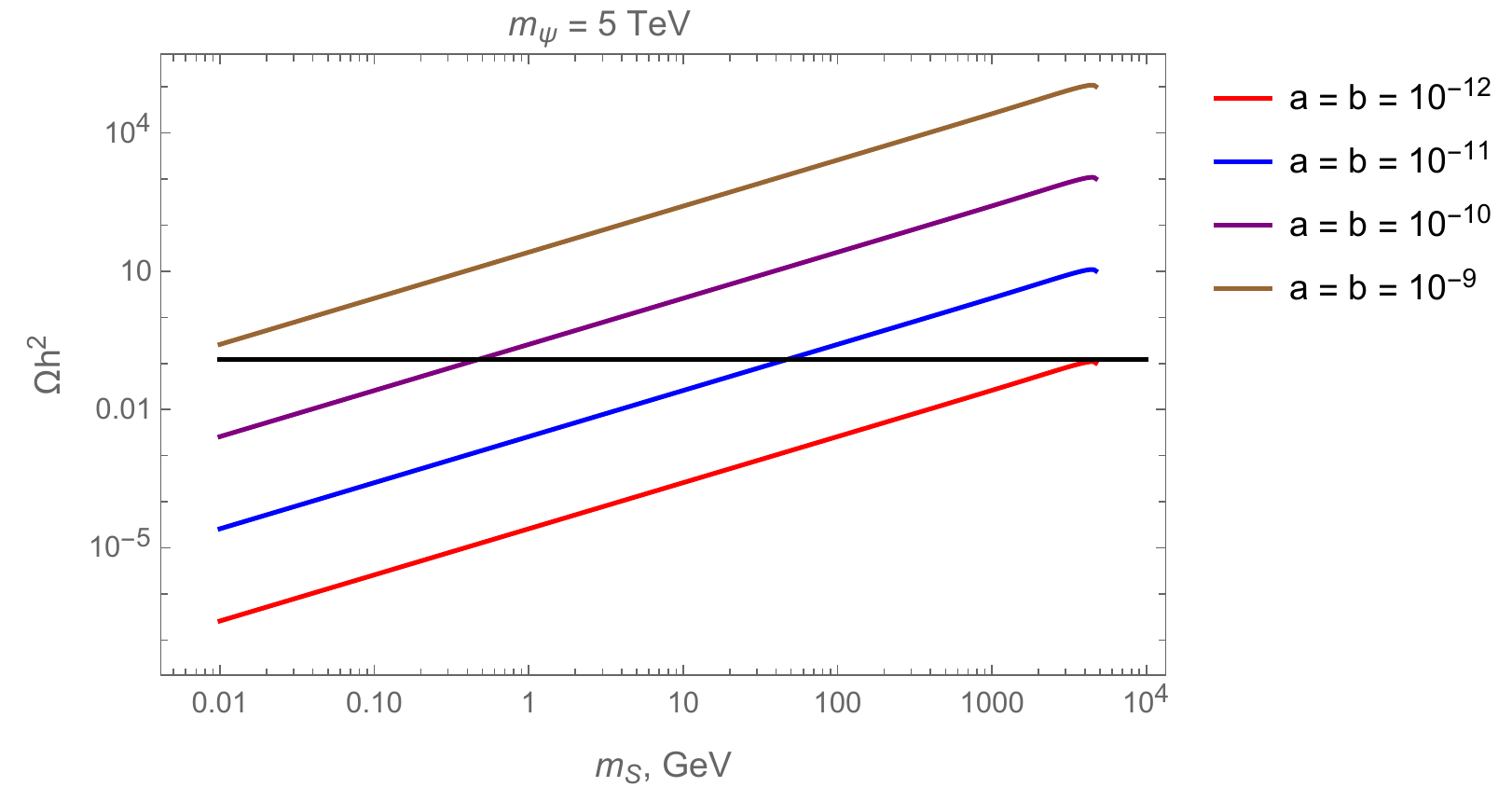}\hspace{0cm}\includegraphics[width=8.5cm]{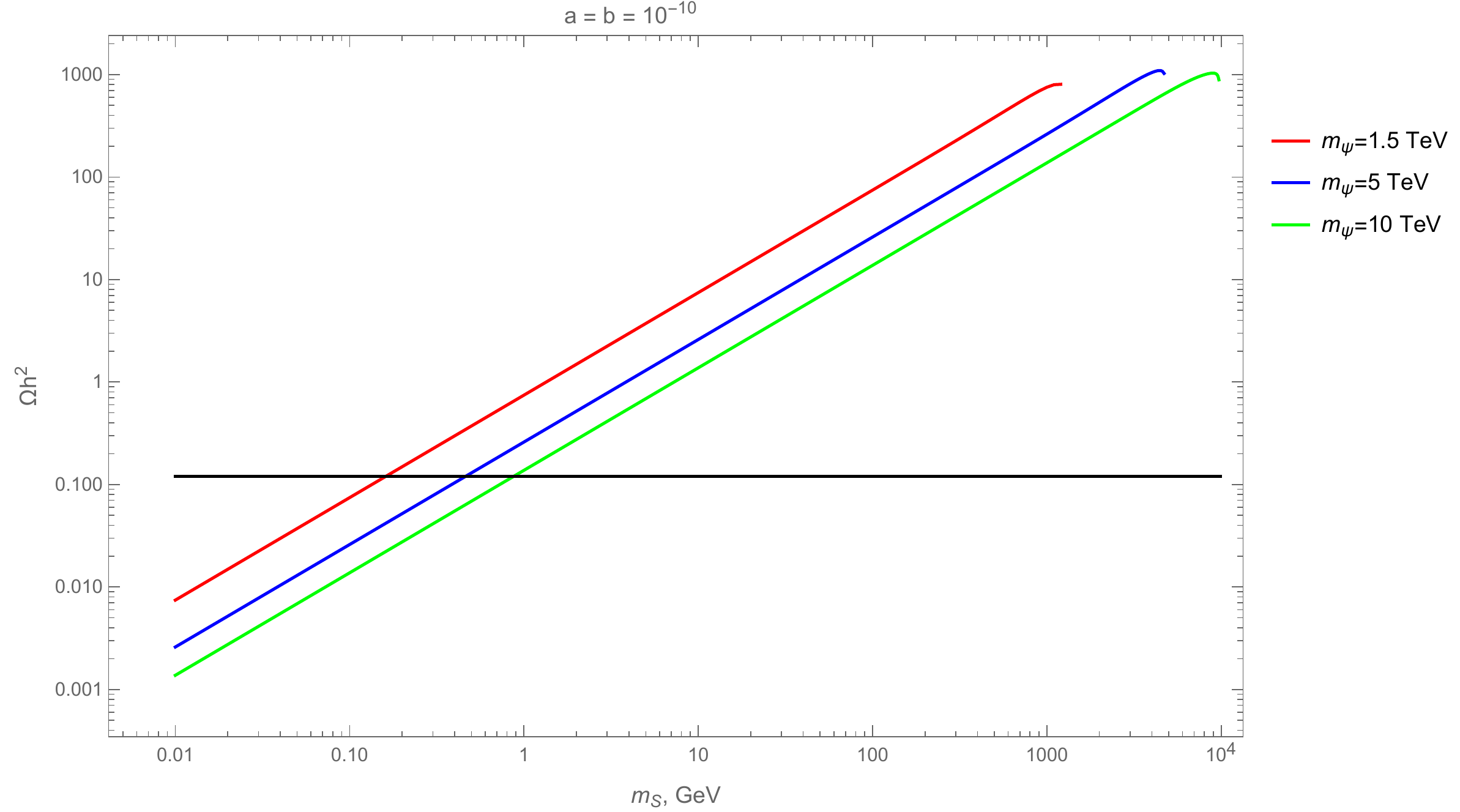}}
\caption{ (Left) The relic abundance of the FIMP dark matter, $S$ with respect to its mass, $m_{S}$ for different values of the Yukawa coupling, $f_{S}=a+i\, b$ and fixed $m_{\psi}$ value, and (Right) the relic abundance of $S$ for different values of $m_{\psi}$ and fixed $f_{S}$ value. Here we consider, $m_{H}>m_{\psi}+m_{S}$ case.}
\label{freezeinfig}
\end{figure}

From Fig. \ref{freezeinfig}, we can see that for $|f_{S}|\sim 10^{-10}-10^{-9}$ and $m_{\psi}\sim O(\mathrm{TeV})$, the FIMP dark matter candidate $S$ can get correct relic abundance for $m_{S}\sim 0.5 - 50$ GeV.

\section{Direct Detection Constraints}
As the DM candidate, $\Sigma^{0}_{R}$ is the component field of the real triplet, $\Sigma_{R}$, it does not have any $Z,Z'$ mediated interactions with the quarks/nucleons. Still it can interact with quarks at one-loop via the exchange of $W^{\pm}_{L,R}$. But the interaction involving $W^{\pm}_{L}$ will be suppressed by $O(\zeta^{2})$ where, $\zeta\sim \frac{m^{2}_{W_{L}}}{m^{2}_{W_{R}}}\frac{m_{b}}{m_{t}}\sim 10^{-5}$. On the other hand, $W^{\pm}_{R}$ exchange will give $O(1/m^{4}_{W_{R}})\sim O(1/\mathrm{TeV}^4)$ mass suppression at the amplitude level. Hence, we can see that the $\Sigma^{0}_{R}$ is still beyond the reach of the current direct detection experiments.

In the case of FIMP dark matter candidate, $S$, its interaction with SM particles only controlled by the Yukawa term Eq. \ref{sint2}. Besides, the interaction with quarks will take place at one-loop level. As the constraint of DM relic density via freeze-in mechanism sets the Yukawa coupling of the order $|f_{S}|\sim O(10^{-10})$, it is also highly suppressed and beyond the reach of current direct detection experiments.

\section{Conclusions} 

In this paper we have studied the simplest left-right model extended to 
include one or more fermion multiplets, such that the latter are predestined 
dark-matter candidates without the imposition of any additional symmetry. 
We studied two examples, one with an $SU(2)_R$ fermion triplet 
$(\Sigma^+_R,\Sigma^0_R,\Sigma^-_R)$, and the other with a fermion singlet 
$S$ and a fermion bidoublet $\psi$.  In each case, the new particles are 
automatically odd under a $Z_2$ symmetry, because of the multiplet structure 
of the model without being imposed.  Hence a stable particle emerges as a 
potential dark-matter candidate.  We analyzed the relic abundance and the 
direct detection prognosis in these two cases, and found the allowed ranges of 
parameter space for each to be a viable dark-matter scenario. 

It is worth noting that having multiple components of dark matter is quite 
natural, as it is questionable that more than $25\%$ of the matter in our 
universe is given by a single particle, while the visible matter, which is 
less than $5\%$ of the matter in the universe is composed by different types 
of quarks and leptons.  If dark matter has two or more components, the 
stringent constraints from relic abundance can be eased or satisfied by 
one of those particles, while the direct detection can be fulfilled by 
another.

\noindent \underline{\it Acknowledgment}~:~
We thank Julio Leite for an important correspondence.  The work of E. M. 
was supported in part by the U.~S.~Department of Energy Grant 
No. DE-SC0008541.  Also, S. K. acknowledges support from Science, 
Technology $\&$ Innovation Funding Authority (STDF) Egypt, under 
grant number 37272.

\end{document}